# *Sticking dynamics of deformable colloids*

Prerna Sharma, Shankar Ghosh and S. Bhattacharya

Department of Condensed Matter Physics and Materials Science

Tata Institute of Fundamental Research

Homi Bhabha Road, Mumbai 400-005, India

## Abstract

The sticking of a soft polystyrene colloidal particle to a planar glass plate was studied by a microrheological technique using an optical tweezer to trap the particle and a piezoelectric-stage to position the plate and to sinusoidally drive it in-planeThe sticking process was well-described by a crossover of the mechanical coupling between the plate and the particle from being dominantly viscous to becoming dominantly elastic. By varying three parameters in the experiment, namely, the strength of the tweezer potential, the strength of the particle-plate interaction and the strength of the oscillatory drive, three different regimes of dynamics - stuck, non-stuck and ageing – were observed, corresponding to the coupling that is elastic, viscous and viscoelastic, respectively. The observations of [1] can be illustrated better by a parametric Nyquist plot of the real and imaginary parts of the effective rigidity modulus of the coupling medium. Upon contact between the particle and the plate, these parameters change abruptly from their bulk values, regardless of whether the particle is stuck or non-stuck in the terminal steady state. In the ageing regime, where sticking evolves gradually in time, the system shows a parametric plot significantly different from what would be expected from a single-relaxation process. A tendency of clustering of the data in the Nyquist plot provides an interesting contrast between truly many-body systems and one with a few degrees of freedom.

# Introduction

A large reduction in the single particle diffusivity or mobility in an interacting many-body system occurs at a variety of transitions, such as (a) the equilibrium freezing transition of a liquid into a crystalline solid, (b) a non-equilibrium freezing of a liquid into a glass, (c) the pinning of a disordered elastic medium in a random potential and (d) the jamming transition in granular matter. Possible relationships among these disparate phenomena are a subject of great current interest and active research [2-4]. Furthermore, the reduction of mobility due to an enhancement of the effective drag is involved in a variety of technologically important but complex processes such as adhesion, lubrication as well as in the general problem of wet and dry friction [5]. It is useful and informative to study a simpler system that contains significant ingredients present in the many body problems of greater complexity but is more tractable in some ways. An example of such a problem is the process of attachment of a soft colloidal particle to a planar glass substrate, which has been investigated recently by a new microrheological technique [1]. Once the particle sticks to the plate due to the attractive Van der Waal attraction, the relative motion between the particle and the plate nearly ceases, resulting in a drastic reduction of the particle's mobility, similar to what occurs in the many-body "phase transitions" described above. The early experimental results [1] showed that the sticking process varies greatly depending upon whether the particle is hard or soft, i.e., whether its motion is governed by the centre-of-mass alone for a hard particle, or if the internal degrees of freedom for a soft particle play an important role. Specifically, the sticking is typically abrupt in the former case but is often gradual in time in the latter. Two different experimental diagnostics were used to quantitatively characterize sticking: (1) the mean squared displacement (MSD) due to the Brownian motion of the particle trapped in an optical tweezer and (2) the phase-sensitive positional response of the particle in the tweezer to the stresses generated by an in-plane oscillatory displacement of the substrate which is a part of a fluid-filled enclosed container. It was shown that the phase difference between the particle and the plate is directly related to the diffusivity of the particle within a Stokes-Einstein formulation of diffusion, using a lumped parameter model based on the centre-of-mass motion of the particle. The amplitude and phase of the motion (or, equivalently, motion of the particle in-phase and in-quadrature) were used to extract a rheological description of the coupling medium between the particle and the plate in terms of the real and imaginary parts of its complex rigidity modulus, G′ and G″, respectively. It was further shown for the soft particle that the coupling medium evolves from a viscous one to an elastic one through an intermediate regime of a time-dependent viscoelasticity within a Maxwellian model. The process is reminiscent of the ageing phenomena in glassy systems. The apparent glassiness was attributed to the internal degrees of freedom, most likely in the form of "polymeric tethers", whose existence has long been speculated upon[6, 7]. The tethers provide a plausible scenario for the observed complex dynamics in this apparently simple system.

In this paper we present both additional measurements and newer ways of presenting the rheological data that further clarify the process of sticking. Additionally, it attempts to address two critical questions: (a) how well the data conform to a simple Maxwellian relaxation model and (b) how one relates the simple quasi-two-body problem studied here to the standard many-body systems mentioned above.

**Experimental methods and Model  :**

The details of the experimental arrangements are given in [1, 8]. A summary is described below and shown in Fig.1. In this paper we focus on the rheological measurements alone. The equation of motion of the center-of-mass of the particle whose displacement, given by x, is:

$$6\pi\eta_{eff}a\left(\dot{x} - \dot{x}_p\right) + k_{opt}x = 0$$

(1)

where $\eta_{eff}$ is the a single lumped parameter viscosity (in principle a complex number) which mediates the stress between the plate and the trapped particle, $k_{opt}$ is the spring constant representing the tweezer potential and $x_p$ is the displacement of the plate. When $x_p$ is varied sinusoidally, i.e., $x_p = x_{p0}$ $e^{i\omega t}$, The solution for x is given by x= $x_0$ $e^{(i\omega t + \phi)}$, where Re(x) and Im(x) are, repectively, $x_0$ cos($\phi$) and $x_0$sin($\phi$). Assuming that the generalized viscosity is $\eta = \eta' + i\eta''$ and further assuming the standard linear viscoelastic notation; $\eta'$ =G''/$\omega$ and $\eta''$ = G'/$\omega$, we obtain the final results for the rheological parameters, rigidity modulus and loss modulus, G' and G'' respectively as:

$$G' = \left(\frac{k_{opt}}{6\pi a}\right)\frac{\text{Re}(x_{res})x_{po} - |x_{res}|^2}{x_{po-}\text{Re}(x_{res})^2 + \text{Im}(x_{res})^2};$$

$$G'' = \left(\frac{k_{opt}}{6\pi a}\right)\frac{\text{Im}(x_{res})x_{po}}{x_{po-}\text{Re}(x_{res})^2 + \text{Im}(x_{res})^2}.$$

(2)

where $x_{res}$ =$x_0$e$^{i\phi}$ and, further,

$$\tan(\phi) = \text{Im}\left(x_{res}\right)/\text{Re}\left(x_{res}\right) = k_{opt}/6\pi\eta a\omega = k_{opt}D/K_BT\omega$$

(3)

where D = $K_BT/6\pi\eta a$ is the Stokes-Einstein diffusion coefficient.

Within this framework, when the particle is far away from the plate, the coupling is expected to be purely viscous for an ordinary fluid, with $\eta'' = 0$ and Im (x) is expected to dominate over Re(x). But once the particle attaches to the plate, all *relative motion* ceases and the response of x is in-phase with that of the plate $x_p$. This implies (a) Im (x) and, therefore tan($\phi$), approach zero and (b) Re (x) increases approaching that of the plate in magnitude, i.e., $x_0 = x_{p0}$. The sticking process is thus described quite simply by the phase, $\phi$. We define an index of sticking s (h,t) as

$$s(h,t) = [1 - \tan \phi(h,t)/ \tan (\phi_0)] = [1 - D(h,t)/D_0 ] \qquad (4)$$

where $\phi_0$ and $D_0$ are, respectively, the phase and the Stokes-Einstein diffusivity of the system while the particle is in the bulk of the fluid where they are nearly independent of the separation h from the plate (h>>a).

## Experimental Results and Analysis

In Figure 2, we describe the typical time-dependent ageing behavior of tan ($\phi$) during the sticking process of the soft polystyrene microsphere. The left half of the abscissa refers to the steady state value of tan ($\phi$) as a function of separation h between the sphere and the plate. Clearly, the polystyrene microsphere takes a very long time (~ 20 minutes) over which the phase decreases jerkily down to the background values. In what follows we focus on the rich variation of the sticking process of the polystyrene microsphere depending on experimentally controlled parameters such as the particle-plate interaction potential U (varied by salt concentration(c)), particle-tweezer interaction potential given by the spring constant $k_{opt}$(varied by the intensity of the laser beam) and the nature of the external drive F (varied by the amplitude and frequency of the external sinusoidal shaking).

Figure 3 shows three different types of behavior of the index of sticking s, as defined in equation (3), observed for different values of the parameters mentioned above, in panels (a) through (c) which we refer to as stuck, non-stuck and ageing, respectively, following the nomenclature of earlier work [1]. In (a) s rises abruptly upon contact to nearly unity just as would be the case silica in Figure 2 and remains very small for all times. In (b), s increases but does not become large enough at long times, which we call unstuck, while in (c), s increases slowly upwards as for the polystyrene case shown in Figure 2 which we call the ageing state.

For each of these cases we compute the rheological parameters G' and G" as in equation (2) and show them in Figure 4, in panels (a) through (c). The three cases can be described by the relative magnitudes of G' and G" in the final steady state. For the stuck state G'>>G"; for the unstuck state the two are comparable but G"> G'. For the ageing state, the system crosses over from G"> G' to G"< G' [9].

In order to relate these observations to the properties of the coupling medium, we note [10] that for conventional liquids and solids one has:

G' = $G_0$ and G" ~ 0                                                    (5a)

for an ideal elastic solid with a rigidity modulus $G_0$ , and

G' ~ 0 and G" = η' ω                                                     (5b)

for a Newtonian liquid of viscosity η'.

These two extremes are interpolated within models of viscoelasticity, the simplest among them being the Maxwell model in which

$$G'(\omega) = G_o \frac{\omega^2 \tau^2}{1 + \omega^2 \tau^2} \text{ a}$$                    (6a)

and

$$G"(\omega) = G_o \frac{\omega \tau}{1 + \omega^2 \tau^2}$$                          (6b)

where τ is a single relaxation time, given by $G_0$ = η'/τ.

A typical situation where the contact is gradual is shown in (c). Note that G" increases upon contact as a function of time, reaches a maximum and then decreases gradually with time. G' starts from a very small value on the left, increasingly monotonically but gradually with increasing time, crossing G" at its peak and then saturates at longer times. As was noted before [1], this behavior is strongly reminiscent of the relaxation behavior of a viscoelastic material as a function of increasing frequency or decreasing temperature in steady state or of an ageing system as at a fixed frequency and temperature as a function of time.

Within the Maxwell model, for example, G' and G' cross at $\omega\tau = 1$ which also coincides with the peak in G". Thus, the time evolution of G' and G in Figure 4(c) can be viewed as a time-dependent enhancement of the effective relaxation time. We note that while the Maxwell model is the simplest conceptual framework for linear viscoelasticity, it seldom works quantitatively for inhomogeneous and highly viscous systems where multiple time scales are usually present. A common generalization of the Maxwell model is achieved by assuming a distribution of relaxation times $g(\tau)$ and that the Maxwell model corresponds to the simplest case of $g(\tau)$ being a delta function $\delta(\tau - \tau_0)$. In our previous study [1], the analogy with a Maxwellian relaxation process was demonstrated by plotting the measured G' and G" against another measured quantity, G'/G", which in the Maxwell model is simply $\omega\tau$. It was observed that the fit to a Maxwellian behavior was good, especially for large values of G'/G". We point out that this agreement is not to be taken as an inevitable validity of the single relaxation time approximation inherent in the Maxwell model. The reasons are as follows.

First, as long as G' reaches a saturation, i.e., G' = a constant and thus G'/G" is proportional to 1/G". At long times, in effect, G" is being plotted against 1/G" and thus one would obtain a slope of -1 in a log-log plot following the peak in G". Similarly, for a Maxwellian, at large values of $\omega\tau$, G" is given in Equation (6b) by G"~ $G_0/\omega\tau$, i.e., proportional to $1/\omega\tau$ . Thus, in effect, $1/\omega\tau$ is being plotted against $\omega\tau$ at large values of $\omega\tau$ and thus a slope of -1 will also obtain. Thus, this is not to be considered to be a validation of the single relaxation time approximation. It does, however, serve as a guide that viscoelasticity is the appropriate framework to describe the time evolution of the rheological property of the stress transmitting medium, namely, the spatially constrained dynamics of the polymeric tethers in the contact region.

A better and more illuminating description of the situation can be obtained by using the generalized Nyquist diagram (better known as the Cole-Cole plot in the case of the Debye model of dielectric relaxation which is formally identical to the Maxwellian viscoelasticity) where the real and imaginary parts of the response function are plotted against each other. To motivate this, we start with the simple freezing of a liquid into a crystalline solid. For the liquid, $G' = 0$ while for a solid $G''/\omega \sim 0$ as in Equation (5). Therefore, a measurement of $G'$ and $G''$ across a freezing transition, say as a function of temperature, would typically collapse into two blobs, one on the y axis for all liquid-state data for which $G''/\omega = \eta$ and the other for all solid-state data on x axis around $G' = G$ with $G'' \sim 0$, on the x-axis, as illustrated in Figure 5(a). On the other hand, if the data represented a glass transition obeying strictly Maxwellian relaxation then the parametric plot would be given by a semi-circle centered at $(G_0/2, 0)$ with a radius of $G_0/2$:

$$\left(G' - G_0/2\right)^2 + G''^2 = \left(G_0/2\right)^2$$

(8)

As the frequency increases at a given temperature, or the relaxation time increases as the temperature is decreased, for example, the system moves from the left corner from a low viscosity liquid through a highly viscoelastic regime on the semi-circle towards a nearly solid regime to the right corner as shown by the arrow in Figure 5(b). The straight line drawn through the middle is the separatrix dividing the dominantly viscous regime and the dominantly elastic regime, going through the (0,0) and the $(G_0/2, G_0/2)$ points on the semicircle. This semi-circle is the characteristic of a Maxwellian or Debye relaxation with a single relaxation time. Most real systems, however, show a marked departure from the semi-circle implying the existence of a distribution of time scales, as mentioned above.

In this context, we analyze our experimental data in Figure 6. Panel (a) where the left panel shows the phase plot in the same manner as before and the right panel shows the resulting Nyquist plot. In (a) the phase dropped sharply at contact but the sticking was abrupt and both G′ and G″ could be measured. Just as in the case of typical liquid-solid freezing, the data collapse into two blobs on the x and y axis with few points in between implying the abruptness of the sticking transition in terms of the rheological properties of the coupling medium going from viscous to elastic. Fig 6(b) shows the other case where the phase decreases significantly but still remains large enough that G″> G′, i.e., the system is still dominantly viscous and remains on the left part of the separatrix separating viscous from elastic. In this case the data once again collapse into blobs, but both representing viscous states, one more viscous than the other. However, the transition is once again abrupt. Fig 6(c) shows the case where the phase decays to very small value gradually, i.e., the so-called ageing case. For this case, the Nyquist plot closely resembles Fig. 5(b). We emphasize that the figure is clearly not a semicircle, i.e., the relaxation is not single valued, nor is the maximum of G″ approximately one-half of the maximum of G′, but rather more akin to Cole-Davidson, Havliriak-Negami or other common empirical forms that represent multiple time scales in a dielectric relaxation process. As stated before [1], this sticking behavior for polystyrene spheres is due to the complex dynamics of polystyrene tethers in the restricted and confined geometry as the plate approaches the particle. Clearly, one needs to understand the differences between a microsphere with internal degrees of freedom and truly many-body systems. In what follows we present some speculative answers to these questions.

Furthermore, Figures 5 and 6 clearly show that the Nyquist plots typically consist of clusters of data points, approximately circumscribed by the blobs drawn in the figures. In all experiments, especially in the ageing cases where one expects much more continuity in the parametric plots, clusters of data points separated by sparse regions are ubiquitous. This is further illustrated in Figure 7 where Nyquist plots of two different runs with different particles are shown. These data as well as the data in Figure 6(c) show commonality with the ideal Cole-Cole plots in Figure 5(b) even though there are significant deviations in the shapes of the curves. But in all cases, the data tend to occur in clusters; even in data containing hundreds of points, one sees only a handful of dense clusters. . Evidence of such clustering can also be seen in the jerky behavior of $\tan(\phi)$ as a function of time. It is tempting to speculate that these clusters reflect the mesoscopic effects present in the system, namely, the existence of only a few basins of metastable minima accessible to the system during sticking. corresponding to specific conformations of the polymeric tethers. The system relaxes among these few minima rather than an extremely large number of them in a truly many body system. Such observations are found in evolutionary biological systems showing punctuated equilibrium [11]. Clearly experimental studies which relate structural characterization of the state of the polymer to the rheological studies are needed to test this hypothesis.

To conclude, we have presented a more instructive analysis of the microrheological measurement of the sticking transition [1]. Using a Nyquist plot, we demonstrate that sticking occurs in three quantitatively different ways in terms of the property of the intervening stress-transmitting medium. The transition is abrupt not only for the sticking case of viscous-to-elastic but also for the non-sticking case of viscous-to-viscous. For the Ageing system, the transition is gradual, well described a relaxational model as shown in the Nyquist plots. But the strong departure from a semi-circle in the parametric plot implies clearly that the system is not well-described by a single relaxation time. However, the concept of viscoelasticity appears to provide intuitive understanding of processes in this system with few degrees of freedom. The situation has close analogy with, for example, the phenomenon of phase organization where simple examples [12] with few degrees of freedom captured using features of the many-body problems [13-15].

Finally, we note that when the drive itself takes the system away from a steady state and nonlinearities such as inherent in memory effects are present [1], the system cannot be described by a simple linear viscoelasticity adopted in the analysis above. The concept of plasticity may be necessary to describe and analyze those cases. Interestingly, that also leads to a closer analogy of the sticking problem with the pinning-depinning transition in disordered elastic media where the transition is generically marked by a wide region of plastic flow above depinning [16, 17].

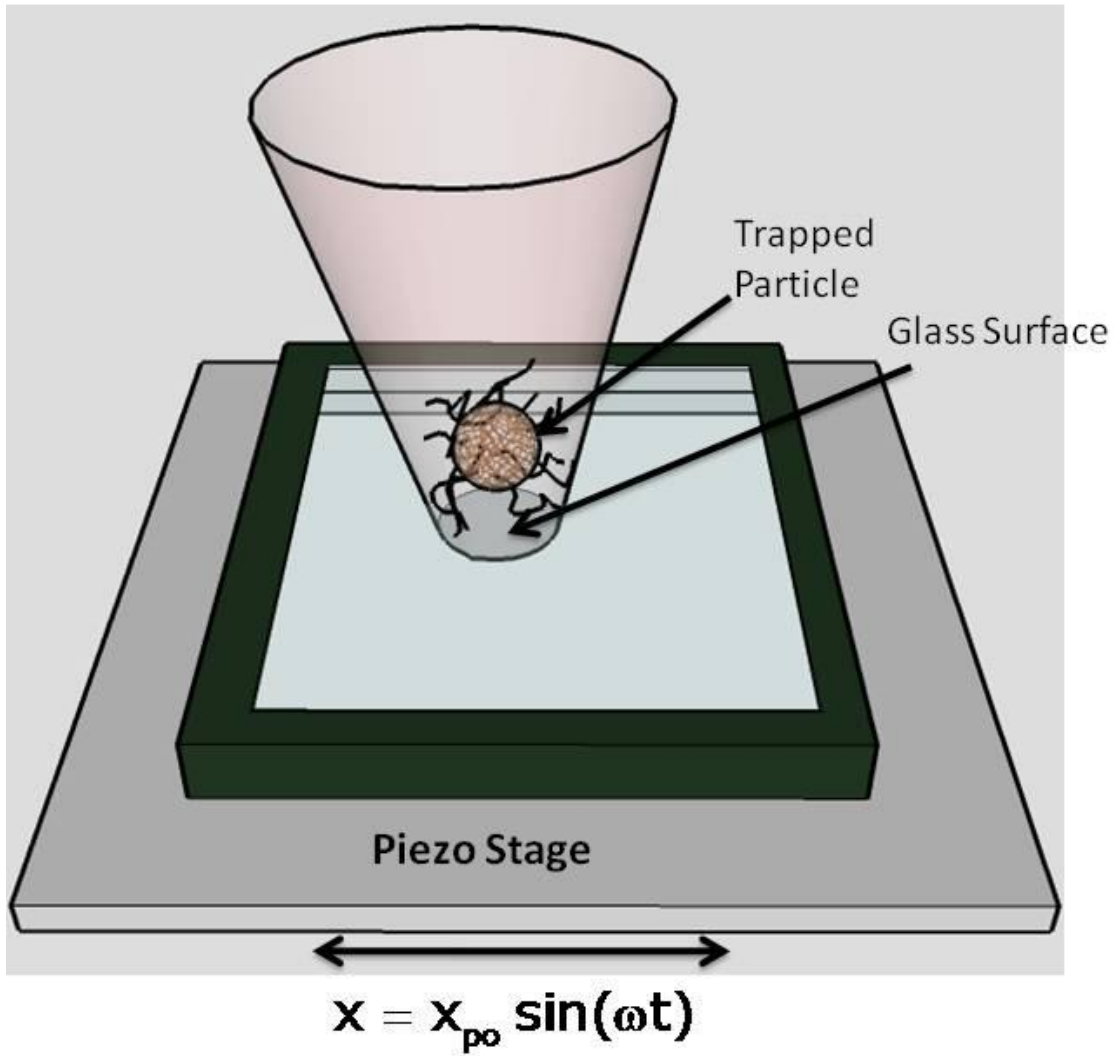

*Figure 1: A schematic of the experimental set-up. The sample cell was made by sticking a rubber o-ring on a microscope cover slip (glass surface) and filled with a dilute colloidal suspension. The components of the optical trapping set-up are the following-(a) trapping laser: Nd-Yag Laser 1064nm. (b) Particle tracking laser: He-Ne laser 632nm. (c) Quadrant Photodiode with a 1KHz bandwidth amplifier. (d) SR830 DSP Lock-In Amplifier. (e) A three-axis Piezo stage from PI Instruments. The details of the experimental set up are given in [1, 8]*

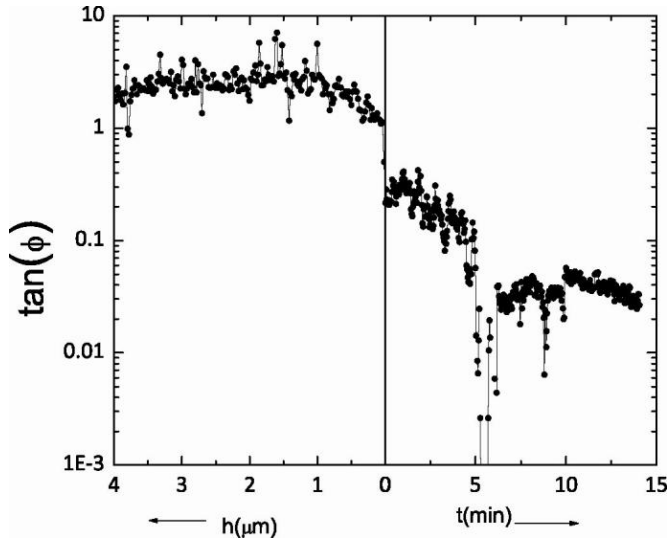

**Figure 2:** *tan($\phi$) of an optically trapped particle in a sealed top container that was oscillated at 251 rad s$^{-1}$ with an amplitude of 33nm in x-direction. The data shown as black circles is for polystyrene microspheres in a 20mM NaCl solution. As h of the particle is decreased by moving the container upwards, towards the particle using the piezo-stage, tan($\phi$)decreases as shown. As tan ($\phi$) = $k_{opt}$/ 6$\pi\eta a \omega$= $k_{opt}D/K_BT\omega$, reaches a value lower than one third of its value at bulk, that is D becomes one-third of its bulk value, we stop the upward motion of the piezo and measure tan($\phi$)when the particle makes a transition from being partly stuck to the glass plate to being completely stuck.(that is, $\phi$ goes to $0^0$ as function of time).*

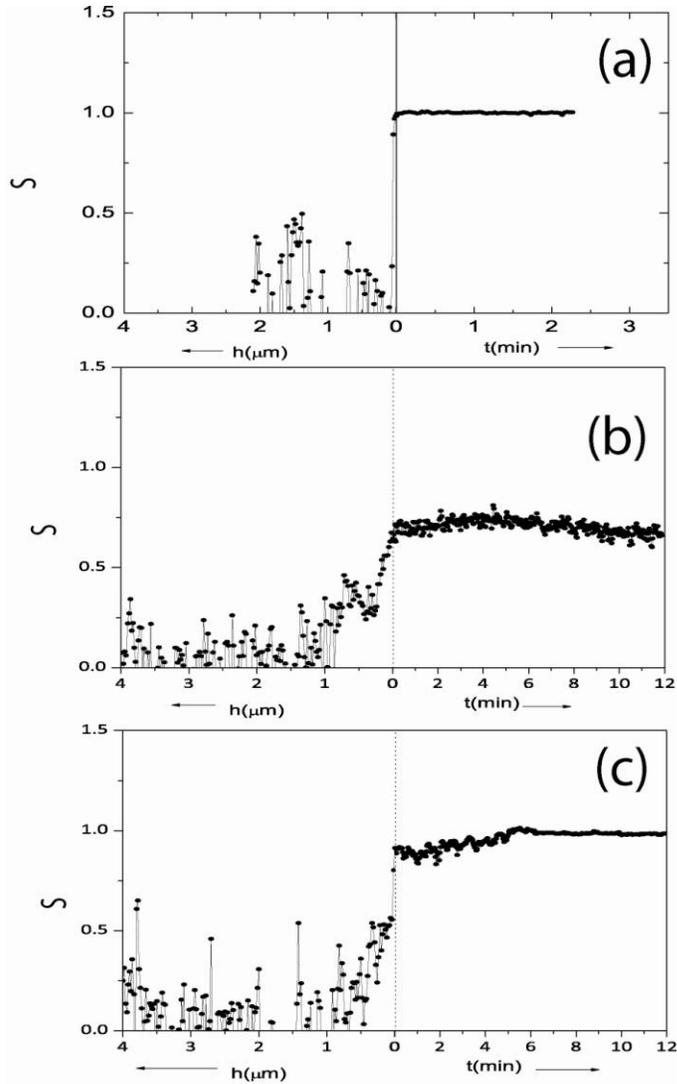

**Figure 3: Sticking strength (s) for (a )Stuck state for the experimental parameters c=30mM, $k_{opt}$=50µNm⁻¹, ω=251 rad s⁻¹ (b) Non-stuck state for the experimental parameters c=10mM, $k_{opt}$t=10µNm⁻¹, ω=251 rad s⁻¹ (c) Ageing state for the experimental parameters c=20mM, kopt=10µNm⁻¹, ω=251 rad s⁻¹ .**

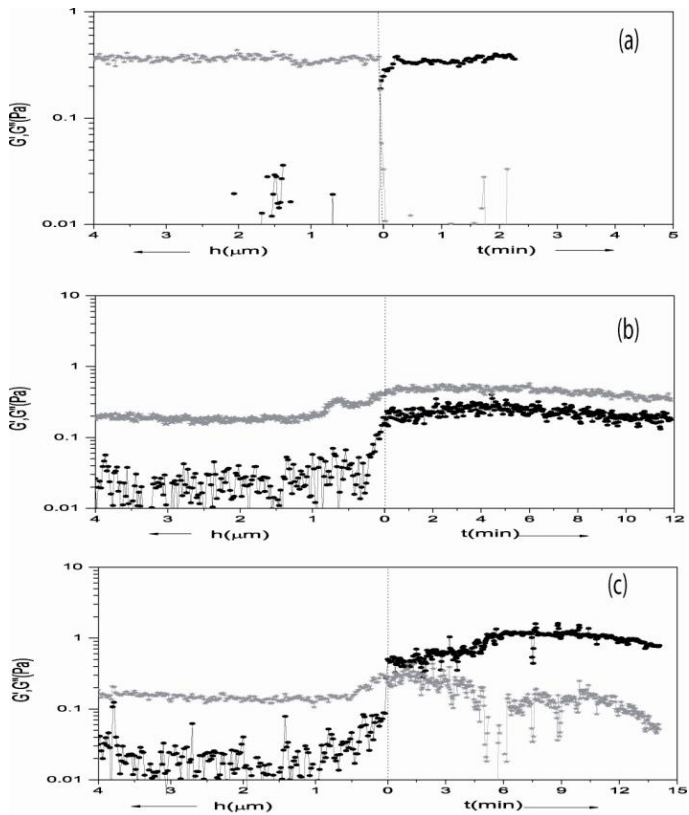

***Figure 4 :Loss modulus (G'') and storage modulus (G') for (a)Stuck state (b)Non-stuck (c) Ageing state  corresponding to the same experimental parameters as in Fig3.***

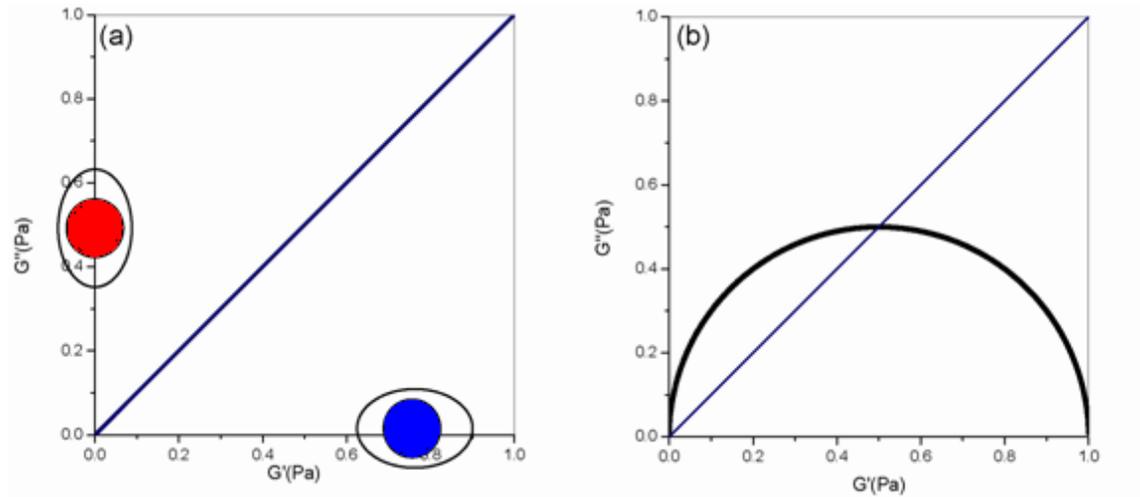

*Figure 5: Cole plots for (a) Liquid (red circle) to solid (blue circle) transition (b)*

*Maxwell process. The straight line (G'=G'') in both the figures notionally divides the*

*liquid state (G''>>G') from the solid state (G''<<G').*

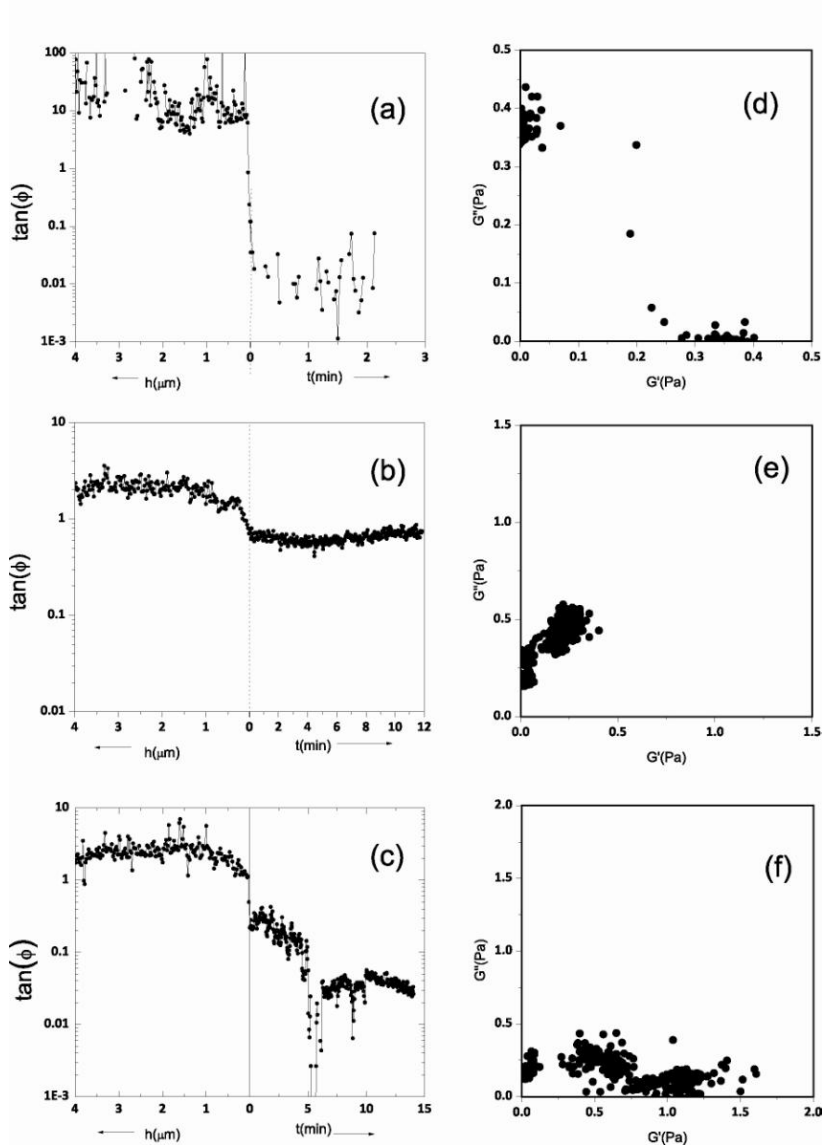

*Figure 6: tan($\phi$) for (a) stuck state (b) non-stuck state (c) aging state. Cole plots for (d) stuck state (e) non stuck state and (f) ageing state. Experimental parameters for these plots were correspondingly the same as in Fig3.*

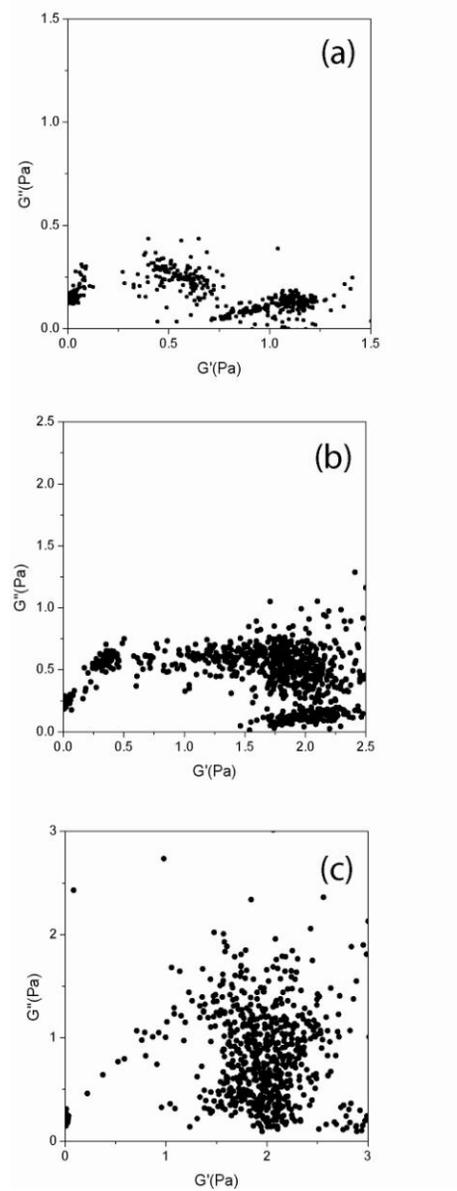

***Figure 7: Cole plots for three ageing states obtained with the experimental parameters
(a) c=20mM, kopt=10μNm⁻¹, ω=251 rad s⁻¹ (b ) c=20mM, kopt=18μNm⁻¹, ω=251 rad s⁻¹
(c) ) c=10mM, kopt=18μNm⁻¹, ω=251 rad s⁻¹.***